\documentclass{aastex61}

\newcommand\aastex{AAS\TeX}

\submitjournal{PASP} 
\shorttitle{\aastex\ sample article}
\shortauthors{Huang et al.}
\begin{document}

\title{Robust Automated Photometry Pipeline for Blurred Images}

\correspondingauthor{Feng WANG}
\email{fengwang@gzhu.edu.cn}

\author{Weirong HUANG}
\affiliation{Center For Astrophysics, Guangzhou University, Guangzhou, 510006, China}
\affiliation{Astronomy Science and Technology Research Laboratory of Department of Education of Guangdong Province, Guangzhou, 510006, China}
\affiliation{Key Laboratory for Astronomical Observation and Technology of Guangzhou, Guangzhou, 510006, China}

\author{Zhou XIE}
\affiliation{Center For Astrophysics, Guangzhou University, Guangzhou, 510006, China}
\affiliation{Astronomy Science and Technology Research Laboratory of Department of Education of Guangdong Province, Guangzhou, 510006, China}
\affiliation{Key Laboratory for Astronomical Observation and Technology of Guangzhou, Guangzhou, 510006, China}

\author{Wenjie ZHONG}
\affiliation{Center For Astrophysics, Guangzhou University, Guangzhou, 510006, China}
\affiliation{Astronomy Science and Technology Research Laboratory of Department of Education of Guangdong Province, Guangzhou, 510006, China}
\affiliation{Key Laboratory for Astronomical Observation and Technology of Guangzhou, Guangzhou, 510006, China}

\author{Ying MEI}
\affiliation{Center For Astrophysics, Guangzhou University, Guangzhou, 510006, China}
\affiliation{Astronomy Science and Technology Research Laboratory of Department of Education of Guangdong Province, Guangzhou, 510006, China}
\affiliation{Key Laboratory for Astronomical Observation and Technology of Guangzhou, Guangzhou, 510006, China}

\author{Hui DENG}
\affiliation{Center For Astrophysics, Guangzhou University, Guangzhou, 510006, China}
\affiliation{Astronomy Science and Technology Research Laboratory of Department of Education of Guangdong Province, Guangzhou, 510006, China}
\affiliation{Key Laboratory for Astronomical Observation and Technology of Guangzhou, Guangzhou, 510006, China}

\author{Yingbo LIU}
\affiliation{Kunming University Of Science And Technology, Kunming, 650504, China}

\author{Feng WANG}
\affiliation{Center For Astrophysics, Guangzhou University, Guangzhou, 510006, China}
\affiliation{Astronomy Science and Technology Research Laboratory of Department of Education of Guangdong Province, Guangzhou, 510006, China}
\affiliation{Key Laboratory for Astronomical Observation and Technology of Guangzhou, Guangzhou, 510006, China}

\begin{abstract}

The primary task of the 1.26-m telescope jointly operated by the National Astronomical Observatory and Guangzhou University is photometric observations of the g, r, and i bands. A data processing pipeline system was set up with mature software packages, such as IRAF, SExtractor, and SCAMP, to process approximately 5 GB of observational data automatically every day. However, the success ratio was significantly reduced when processing blurred images owing to telescope tracking error; this, in turn, significantly constrained the output of the telescope. We propose a robust automated photometric pipeline (RAPP) software that can correctly process blurred images. Two key techniques are presented in detail: blurred star enhancement and robust image matching. A series of tests proved that RAPP not only achieves a photometric success ratio and precision comparable to those of IRAF but also significantly reduces the data processing load and improves the efficiency.

\end{abstract}


\section{Introduction} \label{sec:intro}

The 1.26-m optical/infrared telescope at Xinglong observational station is an equatorial astronomical telescope operated jointly by the National Astronomical Observatory and Guangzhou University in China. It is equipped with a TRIPOL-5 three-channel imaging system and is supported by three charge-coupled device (CCD) cameras (SBIG-STT8300M) to acquire images in the g, r, and i bands simultaneously. Each CDD camera has a resolution of $2504\times3320$ pixels, and the field of view (FOV) is approximately $4.5’\times6’$. Since the telescope began its routine observations, monitoring the short-term variability of blazar objects in the g, r, and i bands simultaneously has been one of its most important tasks. In general, 30–100 photometric observations with exposure times of 300 s should be performed for one blazar object. A daily 8-hour observation can collect nearly 300 CCD images, with a total size of 5 GB.

An important part of state-of-the-art astronomical telescopes is the automation of data processing. To improve the data processing performance and realize scientific production as soon as possible, the telescope has an automated data processing pipeline set up in addition to manual data processing with Image Reduction and Analysis Facility (IRAF) software (\citealt{massey1992user}). Referring to other pipeline systems (\citealt{O'Tuairisgetal2004, Ferreroetal2010, Crossetal2014, Mommert2017, Ma2018, Broutetal2018}), we deployed an automated data processing pipeline (\citealt{Fanetal2019}) that mainly depends on two open-source and mature software packages, i.e., Source Extractor (SExtractor) (\citealt{BertinArnouts1996}) and Software for Calibrating AstroMetry and Photometry (SCAMP) (\citealt{Bertinetal2006}). However, after approximately one year of practical application, the automated data processing system has encountered critical challenges, and many scientists have complained about it. The performance of the automated data processing system strongly depends on the quality of the observational data. In normal situations, the system processes data with a success rate of nearly 95\%. However, for certain types of data, the success rate of the pipeline can be as low as 20\%, which greatly constrains the relevant scientific research.

Analyses have shown that the low photometric success rate is probably attributable to poor image quality and poor tracking performance. However, the telescope accumulates a large amount of observational data when problems are realized. To make full use of the existing data, the algorithm must be optimized effectively to improve the image quality, thereby improving the automatic photometric success rate.

This study develops a more common and robust automatic data processing system for the 1.26-m telescope. Section 2 investigates the drawbacks of existing data processing pipelines. Section 3 presents the key techniques for automating the photometry of blurred images. Section 4 presents the implementation of our robust, star-catalog-independent, platform-independent, and fully automated RAPP as well as the test results. Finally, Section 5 presents the conclusions of this study.

\begin{figure}[ht!] 
\centering
\includegraphics[width=13.0cm, angle=0]{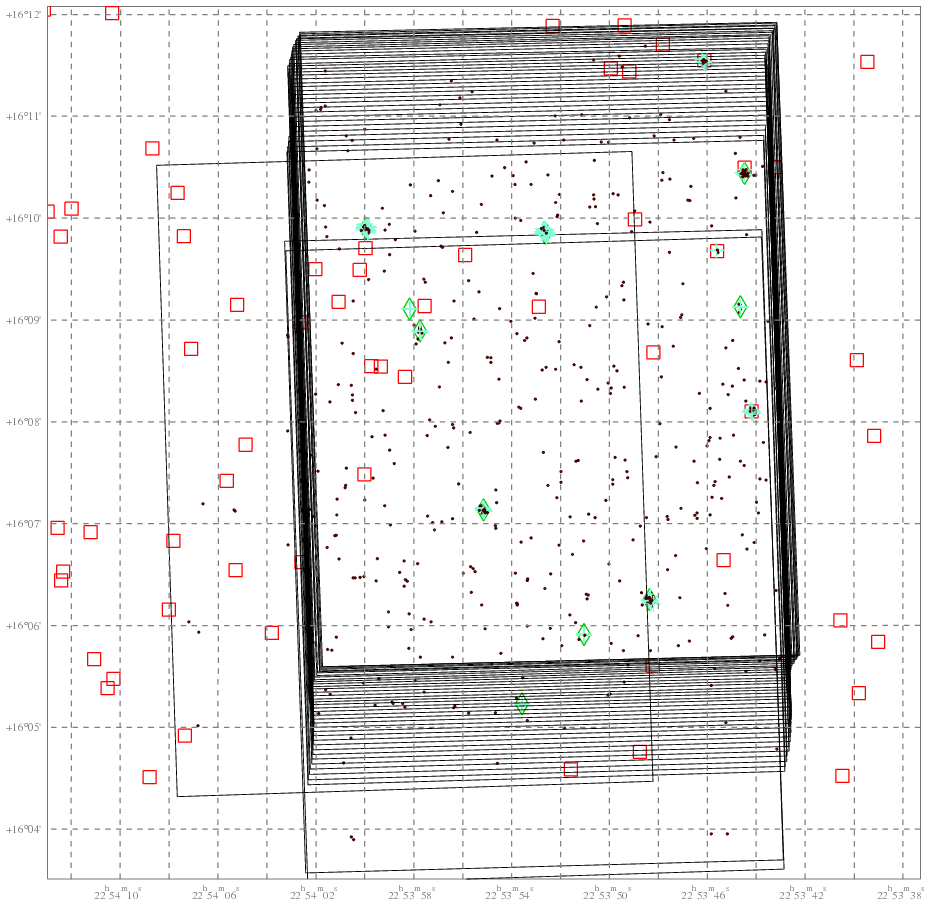}
\caption{Image registration results for data from the 1.26-m telescope processed by the pipeline from \cite{Fanetal2019}. The abscissa is the right ascension, and the ordinate is the declination. The large overlapping black rectangular frame represents each observed image, and the scattered black spots show the stars identified in each image. The red squares and green diamonds represent positions of stars in the catalog, with green representing matching stars.} 
\label{ref_fig1}
\end{figure} 
\begin{figure}[ht!]
\centering
\includegraphics[width=16.5cm, angle=0]{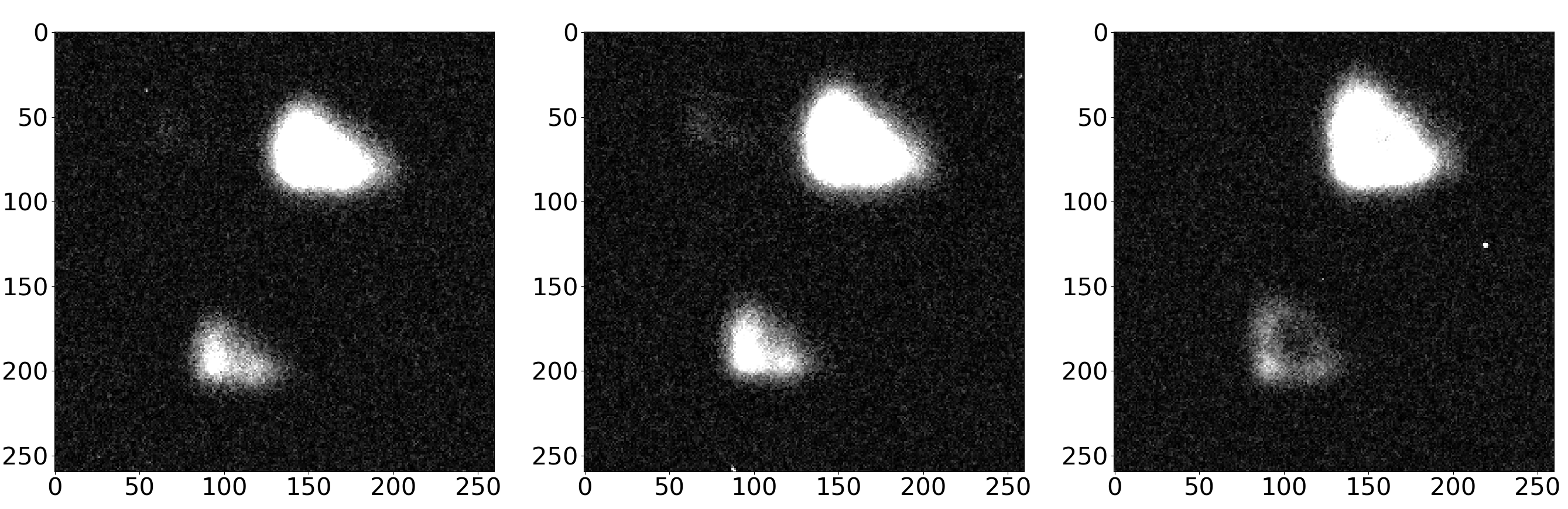}
\caption{
Magnification of partial raw image of blazar 3C454 showing the g, r, and i bands separately, with the same z-scale between $\pm10$ RMS observed by the 1.26-m telescope. The three images were exposed simultaneously, and the target star is in the lower left. This is the most extreme example of problems that are caused by tracking error, showing blur and contour deformity. The influence of poor point-spread-functions on individual exposures and the focus issue may also cause blurring.
}
\label{ref_fig2}
\end{figure}

\section{Analysis of problems with observation data} \label{sec:data}

We systematically investigated all possible reasons that current mature techniques cannot correctly process the telescope’s observational data. We comprehensively analyzed the data and exported detailed intermediate results from these data batch-processing systems, especially those pertaining to star extraction, registration, and aperture photometry radius. The main reasons for the low success rate were image blurring and tracking problems.

To illustrate the impacts of blurred images on the automated photometry processing, we selected a representative data sample observed by the telescope. It was easy to determine that the tracking accuracy of the telescope was insufficient; the pointing direction of the telescope had a large deviation within 4–5 h. Figure~\ref{ref_fig1} shows the image registration results for 50 images generated by SCAMP (\citealt{Bertinetal2006}). The centroid of the FOV showed a significant offset during observation. The first and last images show that the centroid of the FOV shifted by around 1.5’; this is approximately 25\% of the FOV of the telescope. The three images also had abnormal FOV deviations that may have been caused by the observer adjusting the pointing direction of the telescope during the observation or incorrect image registration. The telescope tracking error had a huge impact on the image quality and led to star distortion and a significant decrease in the signal-to-noise ratio (SNR).

The Connected Component Analysis (CCA) algorithm is commonly used; however, its performance strongly depends on the threshold used to extract objects from the background. Both IRAF’s ZPHOT (\citealt{massey1992user}) and SExtractor software (\citealt{BertinArnouts1996}) may use CCA to extract stars using a threshold that is the value of the background plus three times the root mean square (RMS) of the noise.

The poor tracking performance of the telescope leads to image blurring during long exposures of the CCD; this ultimately affects the image quality, especially the brightness and shape of a star. Figure~\ref{ref_fig2} shows three magnified images from the data in the g, r, and i bands. The target star is in the lower left. Compared to the left and middle panels, the shape and brightness of stars in the right panel is distorted and changed. These deviations and distortions reduce the accuracy of star extraction.

\section{Key techniques for automatic photometry of blurred images} \label{3_}

To solve the above-mentioned problems, this paper presents two key techniques: a convolution kernel is used for blurred star enhancement, and the maximum probability matching method is used to address tracking problems.

\subsection{Blurred Star Enhancement Based on Convolution} \label{3_1}

We propose enhancing the signal before star extraction using convolution; this solves the problem of CCA by only using the threshold for weak signals. After the enhancement, the blurred stars can be extracted even if the image is not corrected with bias, dark, and flat fields. This approach guarantees that enough stars are found to run the matching process correctly.

\begin{figure}[ht!]
\centering
\includegraphics[width=7.5cm, angle=0]{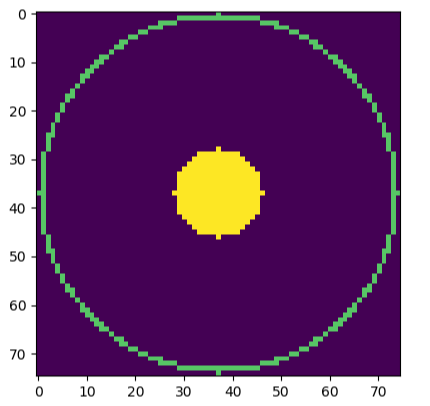}
\caption{
Convolution kernel for blurred star enhancement. The black region has a value of 0, the circular yellow region with the radius $R_{\mathbb{S}}$ has a value of 1, and the annular green region has inner and outer radii $R_{\mathbb{A}}$ and $R_{\mathbb{A}}$+1, respectively, and a value of $-N_{\mathbb{S}}/N_{\mathbb{A}}$
}
\label{ref_fig3}
\end{figure}

For a given image I, assume a circular surface $\mathbb{S}$ with a radius $R_{\mathbb{S}}$ and an annulus $\mathbb{A}$ with inner and outer radii of $R_{\mathbb{A}}$ and $R_{\mathbb{A}}+1$, respectively. Both $\mathbb{S}$ and $\mathbb{A}$ have their centroids at position (x, y). The total flux for the whole $\mathbb{S}$ is equal to the sum of $I$ minus the sky background: $\text{flux}=\sum_{\mathbb{S}}\left(I_{x,y}-B_{x,y}\right)$. The sky background level in $\mathbb{S}$ is approximately equal to the average value in $\mathbb{A}$ ($\overline{B_{\mathbb{A}}}$). Therefore, the total flux in $\mathbb{S}$ is

\begin{equation}
\text{flux}
\approx 
\sum_{\mathbb{S}} I_{x,y}-N_{\mathbb{S}} \overline{B_{\mathbb{A}}}
\end{equation}

where $N_{\mathbb{S}}$ and $N_{\mathbb{A}}$ are the numbers of pixels in $\mathbb{S}$ and $\mathbb{A}$, respectively. When a bright star is blurred, its size becomes bigger, and the value of each pixel ($I_{x,y}$) becomes relatively low or is even drowned out by the sky’s background noise. Therefore, CCA cannot detect it. However, the total flux of a star is always large, and it can be used to detect it effectively. We designed the functional convolution kernel shown in Figure~\ref{ref_fig3} to transfer the image I to flux space ($I_{\text{E}}$). Then, the threshold for determining whether a signal is at the position (x, y) is set to 3 RMS of the sky background in the flux space ($I_{\text{E}}$). The image $I_{\text{E}}$ is given by
\begin{equation}
I_{\text{E}}= h*I 
\end{equation}
where $h$ is the convolution kernel for signal enhancement (Figure~\ref{ref_fig3}).

\begin{figure}
\centering
\includegraphics[width=16.5cm, angle=0]{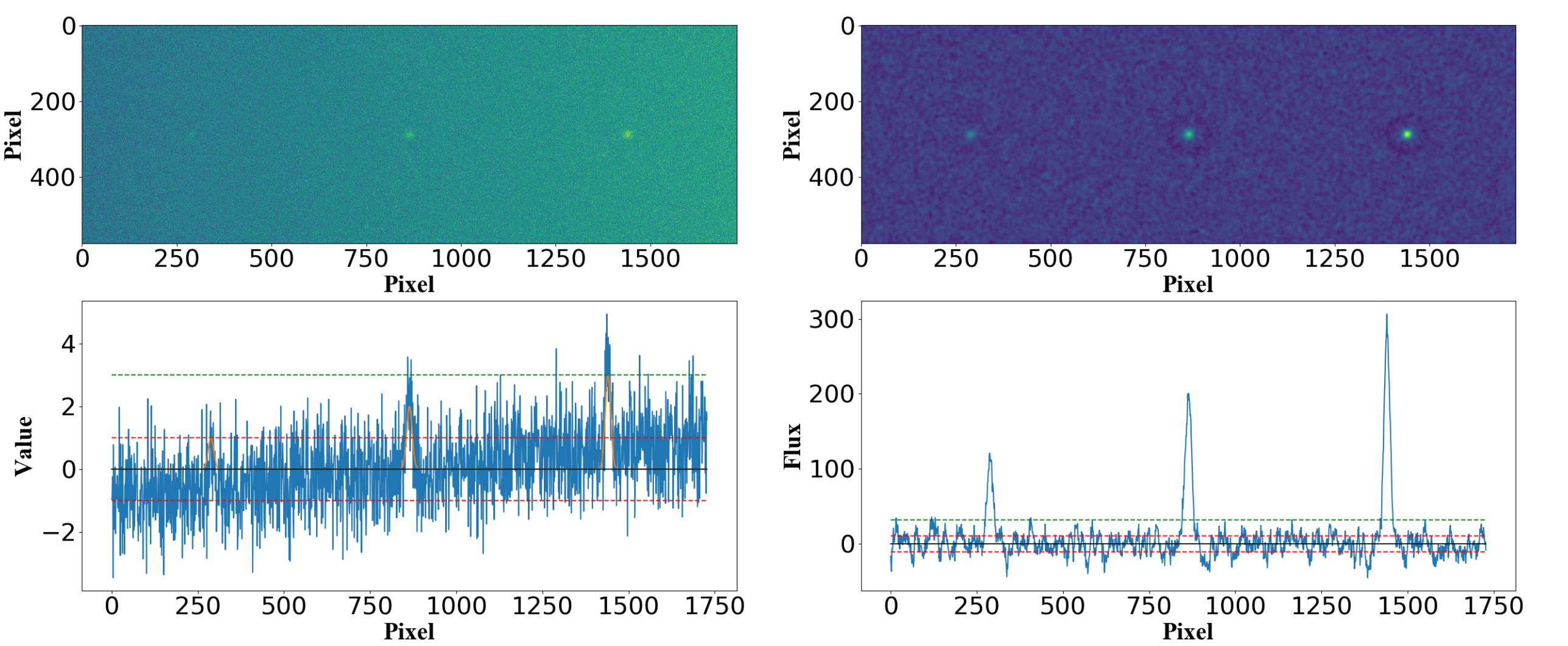}
\caption{
Example of blurred star enhancement. The left image shows a simulation of the observed image ($I$) which includes three blurred stars. The enhanced image on the right ($I_{\text{E}}$) shows that the three stars can easily be recognized after processing with the proposed algorithm. The red dotted lines represent the range of 1 RMS noise, and the green lines represent the range of 3 RMS noise.
}
\label{ref_fig4}
\end{figure}

Figure~\ref{ref_fig4} shows the results of a preliminary test that we performed with this method. The observed image before processing ($I$) and enhanced image after processing ($I_{\text{E}}$) were simulated. Three blurred stars were simulated by the profile of 2D Gaussian (FWHM = 30) in image $I$, and their peak values were 1, 2, and 3 (the SNR of total flux was 17, 35, and 53) respectively. The background of image $I$ is uneven with Gaussian noise (RMS = 1). It shows that the signals that were below 1 RMS noise in $I$ were significantly over 3 RMS noise in $I_{\text{E}}$.

In theory, image $I_{\text{E}}$ yields the maximum contrast when $R_{\mathbb{S}}$ = R$_{\text{star}}$. However, in practice, setting $R_{\mathbb{S}}$ to 9 pixels results in sufficient enhancement. The radius of the annulus is equal to 1.5 times the equivalent radius of the brightest star, which is set to 75 pixels for the 1.26-m telescope. In Figure~\ref{ref_fig4}, the sky background around the star appears slightly sunken, this is caused by the annulus passing through the star. However, this does not affect CCA; it may affect nearby stars in a few cases, but enough stars will remain available for matching. Some connected components extracted by CCA in the enhanced image may not be stars; these are mainly caused by pixels with large values, such as outliers. This detection problem rarely occurs as outliers are filtered before enhancement. In addition, the filtered image is only used to extract stars and not for photometry.

\subsection{Maximum Similarity Matching Method}\label{3_2}

There may not exist many of the bright stars that are used to match images, and two images may contain different stars. For example, the SNR of a star may be reduced, making it undetectable by CCA; different directions of the FOV may result in some stars moving out of one image and other stars moving into another image or connected components in images that are wrongly identified as stars being used for matching. These issues increase the matching difficulty. Therefore, we propose a highly robust method to solve these problems.

\begin{figure}[ht] 
\centering
\includegraphics[width=13.cm, angle=0]{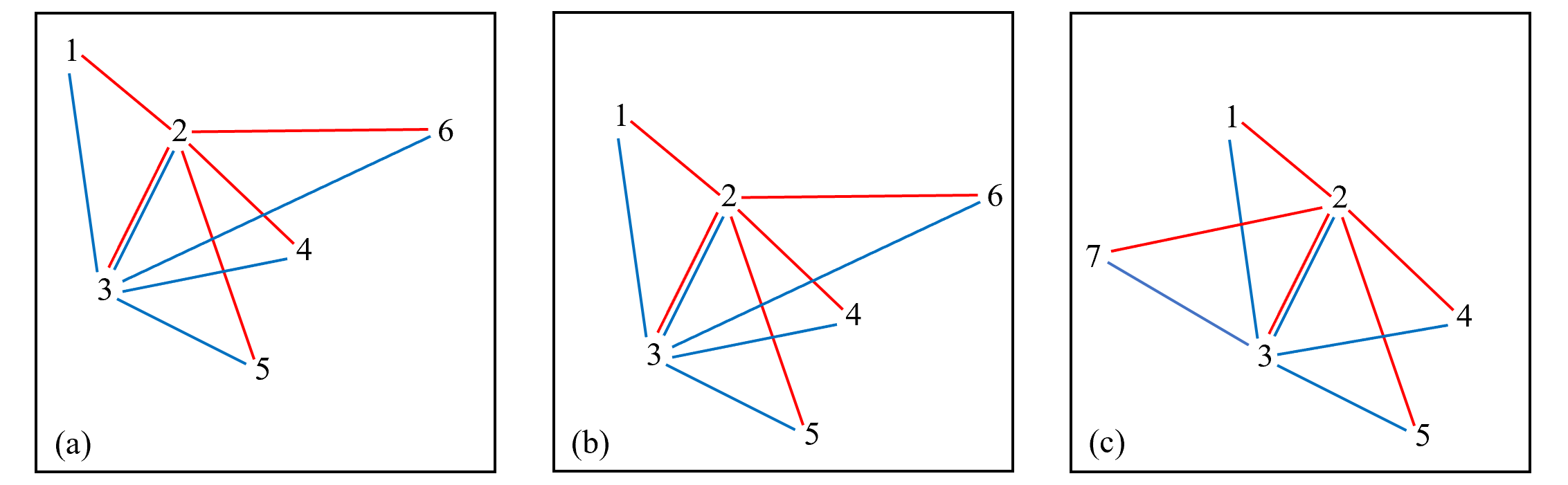}
\caption{
Schematic diagram of the feature sets of the stars (i.e., star 2 and 3). The stars are represented by numbers in the three images. Feature sets $\mathbb{D}_{2}$ and $\mathbb{D}_{3}$ are represented by red and blue lines, respectively. The elements of $\mathbb{D}_{2}$ in different images are mostly the same. However, the elements between D2 and D3 in different images are rarely the same.
}
\label{ref_fig5}
\end{figure}

For the stars in one image, we create a feature set ($\mathbb{D}_{i}$) for each star (see Figure~\ref{ref_fig5}), which is given by
\begin{equation}
\mathbb{D}_{i}=
\left\{ d_{i j} |\  d_{i j}= \left\|\vec{c}_{i}-\vec{c}_{j}\right\|, j \in N
\right\},\ i \in N
\end{equation}
where $N$ is the number of stars in the image, $i$ (or $j$) is the index of star, $\vec{c}$ is the star centroid calculated by $\sum xI_{x,y}/\sum I_{x,y}$ and $\sum yI_{x,y}/\sum I_{x,y}$ (\citealt{Stone1989}), and $d_{i j}$ is the distance between star $i$ and star $j$.

We further define $s(i,i^{\prime})$ as the similarity of stars $i$ in image $I$ ($N$ stars) and star $i'$ in image $I'$ ($N'$ stars), which is expressed by: 
\begin{equation}
s(i, i^{\prime})=
|\mathbb{D}_{i} \cap \mathbb{D}_{i^{\prime}}|,\ i \in N,i' \in N'
\label{equation1}
\end{equation}
where $|\mathbb{D}_{i} \cap \mathbb{D}_{i^{\prime}}|$ is the cardinality of the intersection of $\mathbb{D}_{i}$ and $\mathbb{D}_{i^{\prime}}$, and $i$ and $i'$ are the star indices of the two images. Two elements from $\mathbb{D}_{i}$ and $\mathbb{D}_{i^{\prime}}$ are treated as equal if their difference is less than a certain threshold (1 pixel in this study). Therefore, the value of $s(i, i^{\prime})$ ranges from 0 to $N$ (or $N'$).

Obviously, for star i in image $I$, it is easy to find its corresponding star $i'$ in image $I'$ by locating the maximum value of $s(i,i')$. In theory, at least three star pairs would be searched to calculate the shift distance and rotation angle. 

Having a few different stars in the two images does not impact the matching results (see Figure~\ref{ref_fig5}(a) and Figure~\ref{ref_fig5}(c)). However, in extreme cases, if most of the stars in the two images are different, the failure rate of matching will increase. Therefore, to ensure the robustness of the system, we only use the eight brightest stars for matching. The robustness of the matching algorithm was tested by two simulated images with n stars at fixed locations and 8-n stars at random locations (n = 3, 4, 5). We repeated the simulation 10 million times; Table~\ref{ref_table_FailureRate} shows the results. P(n) is the probability of failure in matching. In practice, the RAPP requires an image with at least three real stars; therefore, the probability of failure is P(3). However, more than three stars can usually be found in each image; therefore, the probability should be P(4) or P(5).

\setcounter{table}{0}
\begin{table}[h!]
\renewcommand{\thetable}{\arabic{table}}
\centering
\caption{Failure rate of matching} \label{ref_table_FailureRate}
\begin{tabular}{cccc}
\tablewidth{0pt}
\hline
\hline
Threshold  & P(3) & P(4) & P(5) \\
\hline
\decimals
0.5  & 5 $\times 10^{-4}$  & 4 $\times 10^{-6}$   & $< 10^{-7}$  \\
1.0  & 2 $\times 10^{-3}$  & 3 $\times 10^{-5}$   & 3 $\times 10^{-7}$  \\
1.5  & 5 $\times 10^{-3}$  & 7 $\times 10^{-5}$   & 2 $\times 10^{-6}$  \\
2.0  & 8 $\times 10^{-3}$  & 2 $\times 10^{-4}$   & 4 $\times 10^{-6}$  \\
\hline
\end{tabular}
\end{table}

\section{system implementation and analysis of results} \label{sec:implementation}
\subsection{Data Processing of the RAPP}

\begin{figure} [htbp] 
\centering
\includegraphics[width=17.5cm, angle=0]{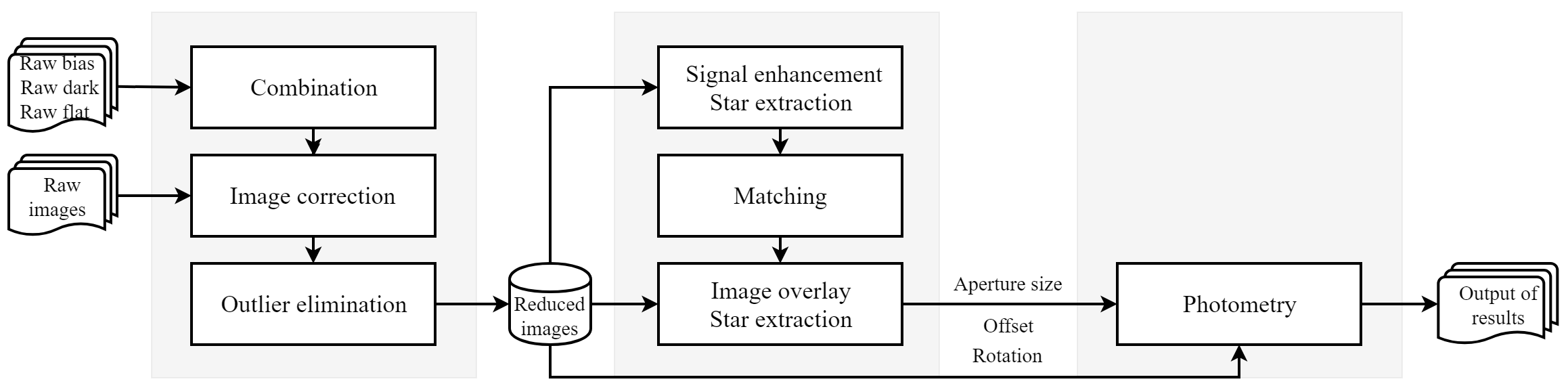}
\caption{Flow diagram of the pipeline. }
\label{ref_fig6}
\end{figure}

The RAPP implemented all main functions for automatic photometry, such as data preprocessing, image matching and overlay, star extraction, and aperture photometry (see Figure~\ref{ref_fig6}).

1)	Data preprocessing. First, bias, dark, and flat field correction was performed on the observational raw data. Meanwhile, the outliers were eliminated using the median of adjacent pixels.

2)	Star extraction and match. Based on the blurred star enhancement technique proposed in Section~\ref{3_1}, the stars were extracted from the enhanced images. And then, we selected the first image as a base image and used the maximum similarity matching method (see section~\ref{3_2}) to align other images. An overlaid image was also generated after the matching of all images.

3) Second star extraction and photometry. The stars were extracted from the overlaid image and aperture photometry \citealt{Costa1992}, the same as IRAF) was performed on the reduced images. We used the same-sized aperture for all stars during photometric processing. The size of the photometric aperture was determined using the maximum diameter of all stars on overlaid image (the users can choose other sizes if needed). 

\subsection{System Implementation and Performance Test} \label{4_1}

We implemented the RAPP based on the techniques discussed in the previous section. The pipeline is written in Python 3, which can be easily implemented and executed in the main system.

\begin{figure} [h]
\centering
\includegraphics[width=17.0cm, angle=0]{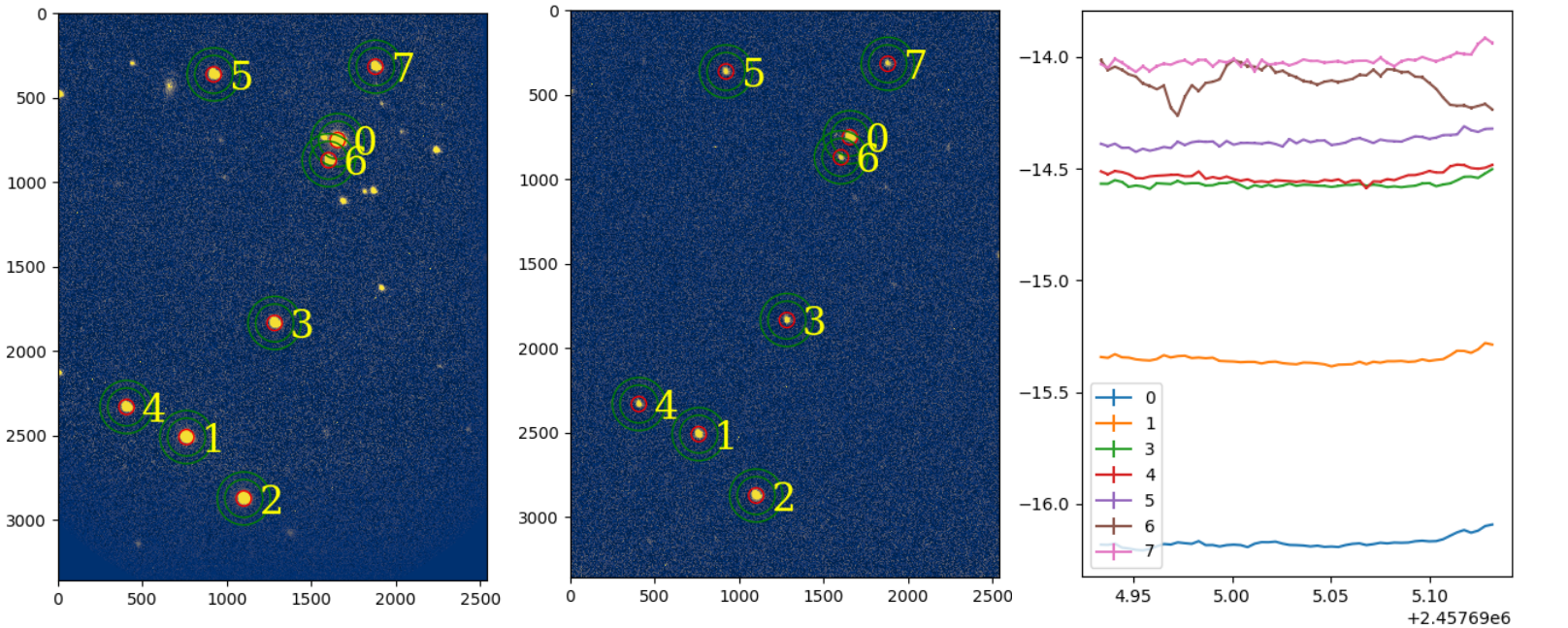}
\caption{
Photometry results in the r band. The data are from observations made on November 2, 2016 and were processed using RAPP. Stars in the images are marked with corresponding numbers and photometric apertures (red circles). The left, middle, and right panels show the overlaid image, target image, and curve of the instrument magnitude of each star, respectively.
} 
\label{ref_fig7}
\end{figure}

The output results of the RAPP show three kinds of images (see Figure~\ref{ref_fig7}): the overlaid image (left panel), the raw images with aperture mark (middle panel), and the light curve of the instrument magnitude for each star (right panel), which is uncalibrated. The magnitude can be easily calibrated further by the users according to their requirements. The output data are archived according to each star’s number in the sequence, and the user can select data by referring to these numbers.

The run time of the automated RAPP was tested. The test laptop was a Dell XPS 13-9360. The operating system was Linux Deepin Version 15.11, and the test environment was Anaconda 3 Python 3.73. Taking the data of Figure~\ref{ref_fig7} as an example, the 57 images were overlaid and aperture photometry was performed on the eight brightest stars (as set by the user). The total run time was 833.3 s (14.6 s per image). The system has been demonstrated to work well in practice. We estimate that it will take around one month to process the previous three years of observational data, whereas manual processing can take one or more years.

\subsection{Analysis of Results}  

\subsubsection{Photometric Success Ratio} 

We selected a total of 1599 1.26-m observational images, discussed in \cite{Fanetal2019}. After removing the images with no stars, 1513 images were left, and these were used for testing. The pipeline in \cite{Fanetal2019} could not detect the target star (3C454.3) in 299 images, whereas RAPP detected all the target stars in 1513 images ($R_{\mathbb{S}}$ of the convolution kernel is 9).

\begin{figure}[h]
\centering
\includegraphics[width=9.cm, angle=0]{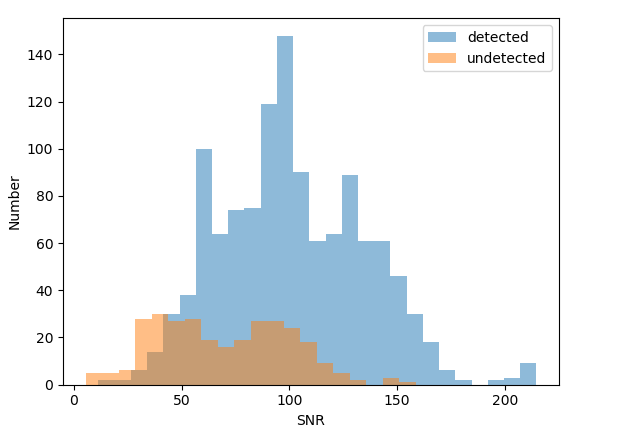} 
\caption{
Histogram for the SNR of the total flux. The blue and orange distributions indicate data for target stars that can and cannot be detected by the previous pipeline, respectively (but can be detected by RAPP). All data can be detected by RAPP.
} 
\label{ref_fig8}
\end{figure}

Figure~\ref{ref_fig8} shows a histogram of the SNR of the total flux of the target star. The blue and orange areas indicate stars that can and cannot be detected by the previous pipeline, respectively. We found that the previous pipeline can detect stars with low SNR of total flux (see blue area in Figure~\ref{ref_fig8}). Analyses suggest that some stars with relatively high SNR of total flux cannot be detected by the previous pipeline, possibly owing to blurred images. In contrast, the RAPP proposed in this paper can detect all the stars (both the blue and orange areas in Figure~\ref{ref_fig8}), thus verifying the robustness of the pipeline.

\subsubsection{Photometric Accuracy Analysis} \label{4_3}

\begin{figure}[h!]
\centering
\includegraphics[width=16.5cm, angle=0]{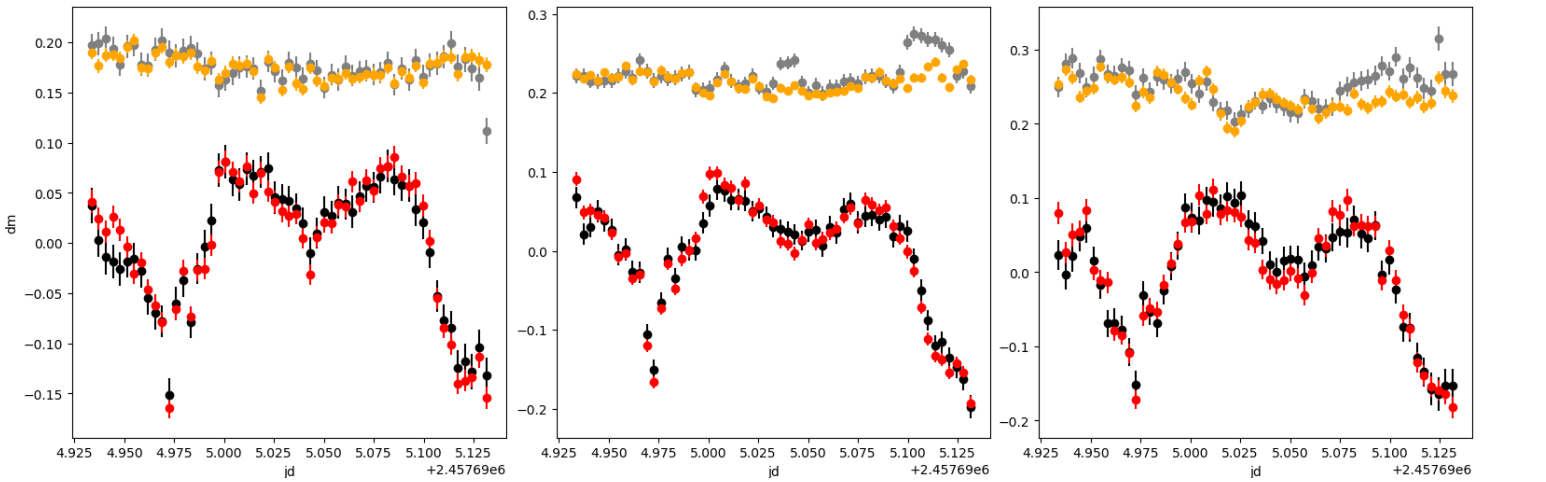}
\caption{
Comparison between the results obtained using RAPP and IRAF. The left, middle, and right panels show the g, r, and i bands, respectively. The red and black dots indicate the result of the target star as obtained using RAPP and IRAF, respectively. The orange and gray dots indicate the result of the reference star as obtained using RAPP and IRAF, respectively.
}
\label{ref_fig9}
\end{figure}

We used the RAPP to process the observational data for 3C454.3 from November 2, 2016 in the g, r, and i bands and compared the results with those of IRAF (APPHOT) from \cite{Fanetal2019}. Figure~\ref{ref_fig9} compares the photometry results in the g, r, and i bands.

Using RAPP, we can get the residual value from the target star’s light curve by subtracting the one obtained using IRAF. The standard deviations of the residuals for the g, r, and i wavebands were 0.015, 0.015, and 0.021, respectively, and the average magnitude errors with IRAF were 0.016, 0.013, and 0.019, respectively. Therefore, the standard deviation of the residual error is close to the magnitude error, showing that the difference between automated processing and IRAF fell within the error range.

For the reference star (in Figure~\ref{ref_fig9}), the light curve with RAPP had standard deviations for the g, r, and i wavebands of 0.011, 0.011, and 0.019, respectively, assuming that the magnitude of the reference star is not time-dependent. With IRAF, the corresponding standard deviations were 0.015, 0.019, and 0.023, respectively. A smaller standard deviation gives a flatter light curve. The results show that RAPP provides a flatter curve than IRAF.

\begin{figure}[h]
\centering
\includegraphics[width=11.5cm, angle=0]{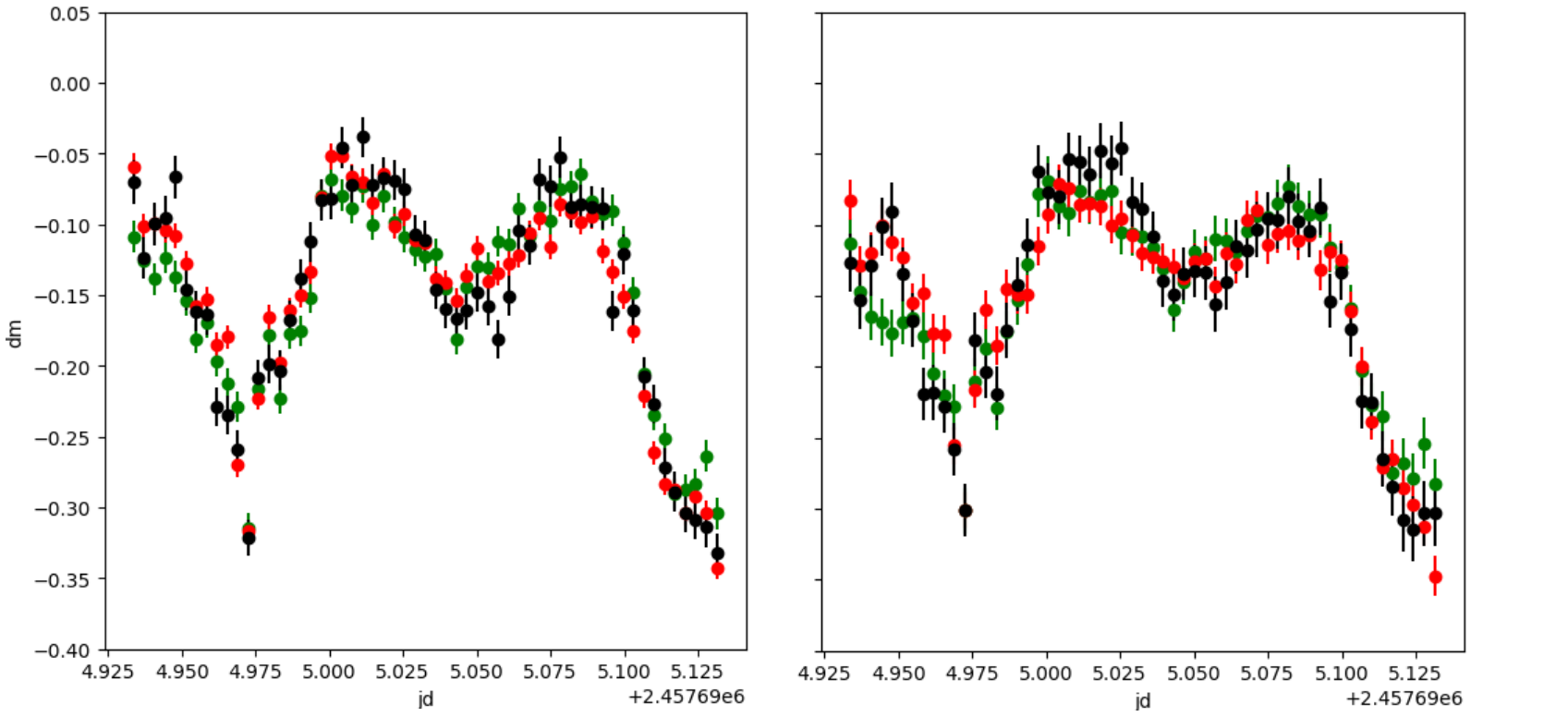}
\caption{
Comparison of the g, r, and i band results. The left and right panels show the results obtained using RAPP and IRAF, respectively. The green, red, and black dots indicate the g, r, and i results, respectively.
}
\label{ref_fig10}
\end{figure}

Figure~\ref{ref_fig10} shows the light curves of the g, r, and i bands in the same panel. The left and right panels show the results of RAPP and IRAF, respectively. We found that the light curve in the three bands as obtained using RAPP shows slightly better consistency.

Overall, RAPP provided almost the same results as did manual processing using IRAF. However, a comparison of the results (see Figure~\ref{ref_fig9}) showed some minor offset between them. Analysis indicated that there are two reasons for these differences:

1)	Different photometric aperture. The RAPP automatically determines the size of the photometric aperture according to the size of the target and reference stars in the overlaid image. However, the IRAF (APPHOT) generally uses the size of the target star as its photometric aperture.

2)	Outliers removal. The RAPP removes outliers in each image because these apparently affect the accuracy of the centroids, and sky background would be affected by these outliers, especially for weaker stars.

\section{Conclusions} \label{sec:conclu}

We proposed and implemented RAPP software that can correctly process blurred images. RAPP requires that the observed images contain at least three stars. Using RAPP greatly reduces the data processing load and improves the efficiency. This paper presents the details of two key techniques for processing blurred images: blurred star enhancement and robust image matching. Tests proved that RAPP is fast and robust while processing point sources of low-quality images. The output results of RAPP showed a noticeably higher photometric success rate than the previous pipeline and had accuracy comparable to that of manual processing with IRAF. The source codes of the pipeline can be downloaded from \url{https://github.com/astronomical-data-processing/RAPP}.

Our pipeline still has some limitations. In its current form, a successful match requires an image with at least three stars for which the SNR of the total flux is greater than $\sim4.5$. The faintest star that can be detected is determined by whether it can be extracted from the overlaid image. However, if only a few observed images are available for a specific day, images observed on different days can be used to operate the image overlay.
 
\acknowledgments
This work is supported by the National Key Research and Development Program of China (2018YFA0404603), the Joint Research Fund in Astronomy (U1831204,U1931141) under cooperative agreement between the National Natural Science Foundation of China (NSFC) and Chinese Academy of Sciences (CAS), the fund for International
Cooperation and Exchange of the National Natural Science Foundation of China (11961141001), the fund of the National Natural Science Foundation of China (11903009,11703010), the Yunnan Key Research and Development Program (2018IA054), the open research program of CAS Key Laboratory of Solar Activity (Nos. KLSA202016), and the major scientific research project of Guangdong regular institutions of higher learning (2017KZDXM062).

\end{document}